\documentstyle[epsfig]{mn}

\title[Galaxy models with power law logarithmic slope]{Spherical galaxy models with power law logarithmic slope}
\author[V.F. Cardone et al.]{V.F. Cardone$^{1}$ \thanks{Corresponding author\,: {\tt winny@na.infn.it}},
E. Piedipalumbo$^{2,3}$, C. Tortora$^2$ \\
$^1$ Dipartimento di Fisica ``E.R. Caianiello'', Universit{\`{a}} di Salerno and INFN, Sezione di Napoli, \\
Gruppo Collegato di Salerno, Via S. Allende, 84081 - Baronissi (Salerno), Italy \\
$^2$ Dipartimento di Scienze Fisiche, Universit\`{a} degli studi di Napoli ``Federico II'', \\
Complesso Universitario di Monte S. Angelo, Via Cinthia, Edificio N, 80126 Napoli, Italy\\
$^3$ INFN, Sezione di Napoli, Complesso Universitario di Monte S. Angelo, Via Cinthia, Edificio G, 80126 Napoli, Italy}

\date{Accepted xxx, Received yyy, in original form zzz}

\pagerange{\pageref{firstpage}--\pageref{lastpage}}
\pubyear{2004}

\begin{document}

\maketitle

\label{firstpage}

\begin{abstract}

We present a new family of spherically symmetric models for the luminous components of elliptical and spiral galaxies and their dark matter haloes. Our starting point is a general expression for the logarithmic slope $\alpha(r) = d\log{\rho}/d\log{r}$ from which most of the cuspy models yet available in literature may be derived. We then dedicate our attention to a particular set of models whose logarithmic slope is a power law function of the radius $r$ investigating in detail their dynamics assuming isotropy in the velocity space. While the basic properties (such as the density profile and the gravitational potential) may be expressed analytically, both the distribution function and the observable quantities (surface brightness and line of sight velocity dispersion) have to be evaluated numerically. We also consider the extension to anisotropic models trying two different parameterization. Since the model recently proposed by Navarro et al. (2004) as the best fit to their sample of numerically simulated haloes belongs to the family we present here, analytical approximations are given for the most useful quantities. 

\end{abstract}


\section{Introduction}

Investigating the kinematics and the dynamics of galactic systems is one of the classical task of astronomy. To complicate the issue, a galaxy is often made out of different subsystems and each one of them has its peculiarities that renders modeling the galaxy a quite difficult task. Along the years, this fascinating challenge has stimulated a lot of astrophysicists to propose a plethora of different models that all share the common feature to be able to reproduce a set of observables suited for the particular kind of system one is interested to. 

The surface brightness is the easiest measurable property of a galaxy and is thus the first evidence a {\it model builder} has to take into account. This explains the large attention devoted to the search for models that, when projected on the plane of the sky, were able to reproduce the observed surface brightness, possibly mimicking the successful $r^{1/4}$ law \cite{deVac48}. Illuminating examples of these kind are the models proposed by Jaffe (1983) and Hernquist (1990). It is worth noting that analytical {\it simplicity} is another desirable property of a model. Hence, while it is possible to deproject both the $r^{1/4}$ law and its generalization, the Sersic $r^{1/n}$ profile \cite{Sersic}, this gives rise to a model \cite{MC02} that is too complicated to be indeed useful. 

On the other hand, the dynamics of a galaxy may hardly be explained by its luminous components only thus claiming for the introduction of a dark halo whose shape, nature and structure are difficult to determine in a unique way. The impressive development in processing speed of modern computers have opened the way to more and more detailed simulations of structures formation that turned out to be a valid help in finding realistic and physically motivated models for the dark halo. Actually, also this way is not free of problems mainly because of the unknown effect of numerical convergence and resolution on the asymptotic behaviour of the density profile.  As a result, different simulations do not always give the same results and, moreover, it is not clear how these latter depend on the software used both to simulate the system and to identify bound structures. While there is a general consensus that relaxed systems exhibit a density profile that is well described by a double power law with outer asymptotic slope $-3$, there is still an open controversy about the value of the inner asymptotic slope with proposed values mainly in the range $\sim 1.0 - 1.5$ \cite{NFW97,TBW97,M98,JS02,P03,Nav04}. On the other hand, a similar controversy has arisen over the question whether such cusps are indeed observed in galaxies (see, e.g., Simon et al. 2003 and references therein). However, on galaxy scale, the effect of baryonic collapse and astrophysical feedback processes (such as supernova explosions) may alter significantly the dark halo structure thus biasing in a complicated way the interpretation of the observations. 

The great variety of models that have been proposed could disorientate giving the impression of a disordered collection of density profiles that are so different from each other that is somewhat surprising that all of them are able to describe the same kind of system. It is thus highly desirable to look for common features of all these models in order to see whether they are indeed different and independent descriptions of the same kind of system or rather they may be seen as particular cases of a more general family. A remarkable step in this direction is represented by the $(\alpha, \beta, \gamma)$\,-\,models proposed by Zhao (1996, 1997). Assuming spherical symmetry\footnote{Without loss of generality, we will only consider spherically symmetric models. Although real galaxies are seldom spherically symmetric, they are indeed axisymmetric, while dark haloes are usually assumed to be spherically symmetric being their true shape unknown. Moreover, introducing an ellipticity in the density profile complicates evaluating the model properties without altering the main results.}, the mass density of the Zhao models is\,:

\begin{displaymath}
\rho(r) = \rho_0 \left ( \frac{r}{r_s} \right )^{-\gamma} \left [ 1 + \left ( \frac{r}{r_s} \right )^{1/\alpha} \right ]^{-\alpha (\beta - \gamma)}
\end{displaymath}
with $r_s$ a scale radius, $\beta$ and $\gamma$ the outer and the inner slope of the density profile respectively and $\alpha$ determining the width of the transition region. It is indeed possible to show that most of the previously proposed galaxy models (both cuspy and cored) are obtained by suitably setting the three parameters $(\alpha, \beta, \gamma)$. Moreover, the Zhao models also forms a complete set for constructing general galaxy models or solving Poisson equation in the non spherical case. 

There is still another possible approach to smooth out the differences among spherically symmetric cuspy models. Qualitatively, we can characterize a model by its asymptotic behaviours for $r \rightarrow 0$ and for $r \rightarrow \infty$. A straightforward way to do this is to assign the logarithmic slope of the density profile, i.e. $\alpha(r) = d\log{\rho}/d\log{r}$. While the usual approach consists in giving the mass density and then evaluating $\alpha(r)$, the inverse way is also possible. We first choose a general expression for the logarithmic slope and then evaluate the corresponding density profile by a simple integration. As we will show later, this method allows to reduce most cupy models to some few classes on the basis of the shape of $\alpha(r)$. Moreover, it is worth stressing that $\alpha(r)$ may be easily estimated by numerical simulations of structure formation and is less affected by systematic effects difficult to control. 

The paper is organized as follows. In Sect.\,2 we introduce a general parameterization of the logarithmic slope $\alpha(r)$ and delineates four classes of models that are worth to be investigated. In particular, we present in this paper a detailed study of the kinematics and dynamics of models with power law logarithmic slope. Sect.\,3 is devoted to the estimate of the basic properties\,: we evaluate the mass density, mass profile, circular velocity, gravitational potential and isotropic velocity dispersion of this class of models. The dynamical properties are fully characterized by the distribution function and the density of states that are estimated in Sect.\,4 assuming isotropy in the velocity space. In Sect.\,5, we evaluate the observable quantities, namely the surface density and luminosity weighted projected velocity dispersion, while, in Sect.\,6, we give off the hypothesis of isotropy and investigate the effect of anisotropy on the radial velocity dispersion and the distribution function. We summarize and conclude in Sect.\,7.

\section{The logarithmic slope}

The usual approach to modeling galactic systems (both the luminous components and the dark haloes) starts from the mass density profile and proceeds by evaluating the main dynamical quantities. The increasing resort to numerical simulations of structure formations has however drawn more attention to the logarithmic slope of the radial profile defined as\,:

\begin{equation}
\alpha(r) \equiv \frac{d \log{\rho}}{d \log{r}} \ .
\label{eq: defslope}
\end{equation}
It is indeed easier to compare different galactic models based on their corresponding logarithmic slope. Moreover, the knowledge of $\alpha(r)$ renders quite immediate to study the asymptotic behaviours (towards the centre and/or the infinity) of the density profile. Furthermore, it seems that this quantity is better constrained than the density profile by numerical simulations.

Motivated by these considerations, it is interesting to look for an expression of $\alpha(r)$ that is as more general as possible in order to build spherical galactic models that may be grouped under the same class by mean of the property that they share the same logarithmic slope. To this aim, a useful proposal for $\alpha(r)$ is\,:

\begin{equation}
\alpha = -\beta \times \frac{1 + a (r/r_s) + b (r/r_s)^2}{1 + c (r/r_s) + d (r/r_s)^2} \times \left ( \frac{r}{r_s} \right )^{\gamma}
\label{eq: genslope}
\end{equation}
with $r_s$ an arbitrary scaling radius. It is possible to show that most of the galaxy models proposed up to now in literature may be obtained by inserting Eq.(\ref{eq: genslope}) into Eq.(\ref{eq: defslope}) and integrating with respect to $r$ with the boundary condition $\rho(r_s) = \rho_s$ after having suitably set the parameters\footnote{In Appendix A we give the values of the parameters $(\beta, a, b, c, d, \gamma)$ for some popular models.} $(\beta, a, b, c, d, \gamma)$. Unfortunately, the result we get by letting completely unspecified the above parameters turns out to be extremely complicated involving the exponential of a combination of hypergeometric and trigonometric functions. As a consequence, it is not surprising that also the simplest dynamical quantities (such as the mass profile) are analytically impossible to derive so that Eq.(\ref{eq: genslope}) is useless for astrophysical applications. We have thus to look for general, but simpler expressions for the logarithmic slope $\alpha(r)$.  

To this aim, let us observe that most of the cuspy models frequently used in literature may be deduced by integrating Eq.(\ref{eq: genslope}) with $b = d = 0$ so that, heron, we will restrict our attention to models having the following expression for the logarithmic slope\,:

\begin{equation}
\alpha(r) = -\beta \times \frac{1 + a (r/r_s)}{1 + c (r/r_s)} \times \left ( \frac{r}{r_s} \right )^{\gamma} \ .
\label{eq: quitegenslope}
\end{equation}
Let us first consider the case $c \ne 0$. Without loss of generality, we may set $c = 1$ thus obtaining\,:

\begin{equation}
\alpha(r) = -\beta \times \frac{1 + a (r/r_s)}{1 + (r/r_s)} \times \left ( \frac{r}{r_s} \right )^{\gamma} \ .
\label{eq: semigenslope}
\end{equation} 
If $c \ne 1$, we may rescale all the results obtained starting from Eq.(\ref{eq: semigenslope}) by replacing\,:

\begin{displaymath}
r_s \rightarrow r_s/c \ , \ \beta \rightarrow \beta/c^{\gamma} \ , \ a \rightarrow a/c \ .
\end{displaymath}  
For $\gamma = 0$, Eq.(\ref{eq: semigenslope}) reduces to\,:

\begin{equation}
\alpha(r) = -\beta \times \frac{1 + a (r/r_s)}{1 + (r/r_s)} \
\label{eq: slopeflos}
\end{equation}
that may be considered as a generalization of a lot of double power law models (see Table\,1 in Appendix A). On the other hand, by setting $a = c = 0$ and $\gamma \ne 0$, we get\,:

\begin{equation}
\alpha(r) = -\beta \times \left ( \frac{r}{r_s} \right )^{\gamma} \ .
\label{eq: preslopepolls}
\end{equation}
For $(\beta, \gamma) = (2, 0.17)$, this reduces to the model recently proposed by Navarro et al. (2004, hereafter N04) on the basis of a set of high resolution numerical simulations of dark matter haloes. Finally, for $c = 0$ and $\gamma \ne 0$, we may set $a = 1$ and obtain a fourth class of models with\,:

\begin{equation}
\alpha(r) = -\beta \times (1 + r/r_s) \times \left ( \frac{r}{r_s} \right )^{\gamma}
\label{eq: linearslope}
\end{equation}
that may be generalized to the case $a \ne 1$ by the following replacements\,:

\begin{displaymath}
r_s \rightarrow r_s/a \ , \ \beta \rightarrow \beta/a^{\gamma} \ .
\end{displaymath}
Summarizing, we have defined four set of models characterized by logarithmic slopes given by Eqs.(\ref{eq: semigenslope}), (\ref{eq: slopeflos}), (\ref{eq: preslopepolls}) and (\ref{eq: linearslope}) respectively. We have checked that, while it is possible to get an analytical expression (at least, in terms of special functions) for the density profile in all four cases, the mass profile is analytical only for models with $\alpha(r)$ given by Eq.(\ref{eq: slopeflos}) and Eq.(\ref{eq: preslopepolls}). Our aim here is to study galaxy models that are both general and analytically amenable so that they may be easily compared to observational data (such as those on the rotation curve). That is why only Eqs.(\ref{eq: slopeflos}) and Eq.(\ref{eq: slopepolls}) are worth to be considered in detail. Moreover, it is easy to show that integrating Eq.(\ref{eq: slopeflos}) gives rise to a class of models that is only a subset of the more general family of the Zhao models. This is partially true also for models with $\alpha(r)$  given by Eq.(\ref{eq: preslopepolls}) that may indeed be obtained as a limiting case. However, the dynamical properties of these particular models have not been investigated in detail by Zhao. Moreover, the result of Navarro et al. \shortcite{Nav04} quoted above is a strong motivation to dedicate much attention to these models which we will refer to in the following as {\it PoLLS} ({\it Power Law Logarithmic Slope}) models.

\section{Basic properties}

The logarithmic slope of the radial density profile for PoLLS models is given by Eq.(\ref{eq: slopepolls}) and is characterized by four parameters, namely the scaling radius $r_s$, the characteristic density $\rho_s$ and the two slope parameters $(\beta, \gamma)$. Actually, it is possible to reduce the number of parameters by redefining the scale radius. To this aim, let us evaluate $r_{-2}$, the radius at which the logarithmic slope equals the value of the isothermal sphere. By solving $\alpha(r_{-2}) = -2$, we easily get\,:

\begin{equation}
r_{-2} = \left ( \frac{2}{\beta} \right )^{1/\gamma} r_s \ .
\label{eq: riso}
\end{equation}
Replacing $r_s$ with $r_{-2}$, Eq.(\ref{eq: slopepolls}) may be written as\,:

\begin{equation}
\alpha(r) = -2 \left ( \frac{r}{r_{-2}} \right )^{\gamma}
\label{eq: slopepolls}
\end{equation}
so that the parameter $\beta$ may indeed be eliminated. It is interesting to look at the asymptotic behaviours of $\alpha(r)$\,:

\begin{displaymath}
\lim_{r \rightarrow 0}{\alpha(r)} = 0 \ ,
\end{displaymath}

\begin{displaymath}
\lim_{r \rightarrow \infty}{\alpha(r)} = - \infty \ .
\end{displaymath}
The logarithmic slope does not diverge in the centre so that the model is not singular which is a nice feature, while $\alpha$ is monotonically decreasing thus suggesting an exponential\,-\,like decrease of the mass density.  Actually, the density profile turns out to be\,:

\begin{equation}
\rho(r) = \rho_{-2} \exp{\left \{ - \frac{2}{\gamma} \left [ \left ( \frac{r}{r_{-2}} \right )^{\gamma} - 1 \right ] \right \}} 
\label{eq: rho}
\end{equation}  
so that $\rho$ remains finite for $r \rightarrow 0$, which is what usually  happens in the case of cored models as, e.g., the non singular isothermal sphere \cite{BT87,HK87}. The exponential decrease of $\rho$ is coherent with the asymptotic limit of $\alpha$ for $r \rightarrow \infty$. As a general remark, note that PoLLS models, loosely speaking, may be considered as a one parameter family with $\gamma$ as ordering parameter and $(r_{-2}, \rho{-2})$ scaling quantities.

A strong constraint on $\gamma$ comes from the evaluation of the mass profile\,:

\begin{displaymath}
M(r) = 4 \pi \int_{0}^{r}{r'^2 \rho(r') dr'} = 4 \pi \rho_{-2} r_{-2}^3 
\int_{0}^{x}{\exp{\left [ - \frac{2}{\gamma} (\xi^{\gamma - 1}) \right ]} d\xi} 
\end{displaymath}
having defined $x \equiv r/r_{-2}$. This integral is defined only for $\gamma > 0$ giving\,:

\begin{figure}
\resizebox{8.5cm}{!}{\includegraphics{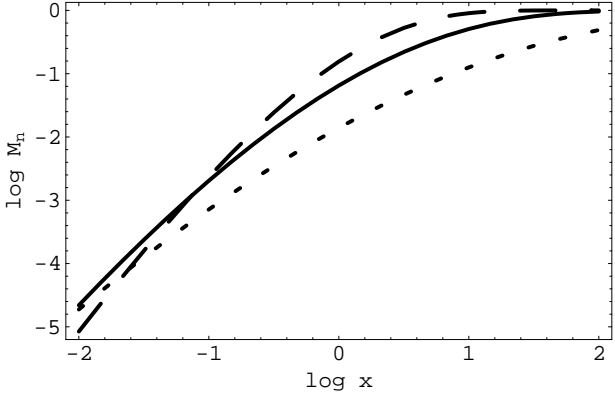}}
\caption{Logarithm of the normalized mass as function of the logarithm of the scaled radius $x$ for PoLLS models with three values of $\gamma$, namely $\gamma = 0.085$ (short dashed), $\gamma = 0.17$ (N04 model, solid) and $\gamma = 0.34$ (long dashed).}
\label{fig: logmass}
\end{figure}

\begin{equation}
M(x) = M_{tot} \times \frac{\Gamma(3/\gamma) - \Gamma(3/\gamma, 2 x^{\gamma}/\gamma)}{\Gamma(3/\gamma)}
\label{eq: mass}
\end{equation}
where $M_{tot}$ is the total mass given by\,:

\begin{equation}
M_{tot} = \frac{4 \pi \rho_{-2} r_{-2}^3 }{\gamma} \left ( \frac{2}{\gamma} \right )^{-3/\gamma} \Gamma(3/\gamma) \exp{(2/\gamma)} \ ,
\label{eq: mtot}
\end{equation}
and $\Gamma(x)$ and $\Gamma(a, x)$ are the usual $\Gamma$ function and the incomplete $\Gamma$ function respectively. In Fig.\,\ref{fig: logmass}, we report the logarithm of $M_n \equiv M(x)/M_{tot}$ for three different choices of $\gamma$. Except in the very inner regions $(\log{x} \le -1)$, for a fixed value of $x$, the higher is $\gamma$, the higher is the ratio among the mass within $x$ and the total one thus indicating that models with higher values of $\gamma$ converges more quickly to the total mass. It is worth noting, however, that $M_n$ converges to 1 only slowly, i.e. for values of $r >> 5 r_{-2}$ for the models we have considered. As a result, the scaled half mass radius $x_{hm}$, implicitly defined as $M(x_{hm})/M_{tot} = 1/2$, turns out to be quite large. For instance, for the N04 model it is $x_{hm} \simeq 10$, while for $\gamma \in (0.06, 1.06)$ $x_{hm}$ is well approximated (within $5\%$) as\,:

\begin{equation}
\log{x_{hm}} \simeq 0.146 \gamma^{-1.066}
\label{eq: xhm}
\end{equation}
as we have checked numerically. As regard the total mass, this is decreasing function of the slope parameter $\gamma$ as it is shown in Fig.\,\ref{fig: totmass}. Qualitatively, this can be explained by noting that models with higher $\gamma$ are more concentrated (i.e. the density decreases faster) so that less mass is contained within a given volume with the result that $M_{tot}$ is smaller.

Being the model spherically symmetric, we do not need to explicitly compute the gravitational potential to evaluate the rotation curve. It is simply\,:

\begin{displaymath}
v_c(r) = \sqrt{GM(r)/r}
\end{displaymath}
that, in our case, reduces to\,:

\begin{equation}
v_c(x) = v_{-2} \times {\cal{V}}(x)
\label{eq: vc}
\end{equation}
having defined\,:

\begin{figure}
\resizebox{8.5cm}{!}{\includegraphics{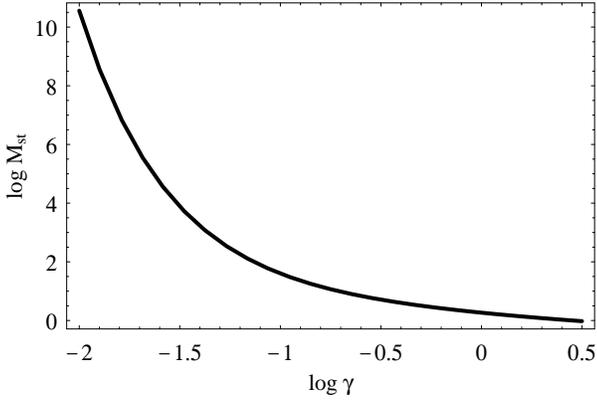}}
\caption{Logarithm of the scaled total mass $M_{st} \equiv M_{tot}/4 \pi \rho_{-2} r_{-2}^3$ as function of the logarithm of the slope $\gamma$.}
\label{fig: totmass}
\end{figure}

\begin{figure}
\resizebox{8.5cm}{!}{\includegraphics{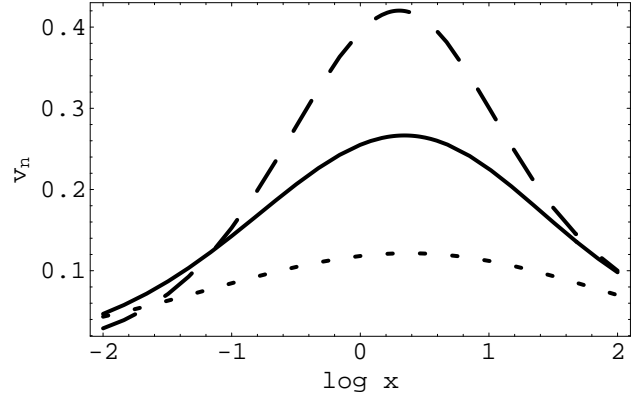}}
\caption{The normalized circular velocity $v_n(x) = v_c(x)/\sqrt{G M_{tot}/r_{-2}}$ as function of the logarithm of the scaled radius $x$ for PoLLS models with three values of $\gamma$, namely $\gamma = 0.085$ (short dashed), $\gamma = 0.17$ (N04 model, solid) and $\gamma = 0.34$ (long dashed).}
\label{fig: vcfig}
\end{figure}

\begin{equation}
v_{-2} \equiv v_c(r = r_{-2}) = \sqrt{\frac{G M_{tot}}{r_{-2}} \left [ 
\frac{\Gamma(3/\gamma) - \Gamma(3/\gamma, 2/\gamma)}{\Gamma(3/\gamma)} \right ]} \ ,
\label{eq: vs}
\end{equation}

\begin{equation}
{\cal{V}}(x) \equiv \sqrt{\frac{1}{x} \times 
\frac{\Gamma(3/\gamma) - \Gamma(3/\gamma, 2 x^{\gamma}/\gamma)}{\Gamma(3/\gamma) - \Gamma(3/\gamma, 2/\gamma)}} \ .
\label{eq: vcfun}
\end{equation}
Looking at Fig.\,\ref{fig: vcfig}, where the normalized normalized velocity  $v_n(x) = v_c(x)/\sqrt{G M_{tot}/r_{-2}}$ is shown, we see that higher values of $\gamma$ gives rise to higher peak velocities. The position $x_{max}$ of the peak is a function of $\gamma$ and is well approximated (within $5\%$) as\,:

\begin{equation}
x_{max} \simeq 1.680 \gamma^{-0.135} . 
\label{eq: xmax}
\end{equation}
For the N04 model, it is $x_{max} \simeq 2.1$, while models with lower $\gamma$ peaks farther from the galaxy centre. The rotation curve is asymptotically keplerian. Indeed, for $x \rightarrow \infty$, $\Gamma(3/\gamma, 2 x^{\gamma}/\gamma) \rightarrow 0$ so that the fractionary term in Eq.(\ref{eq: vcfun}) goes to unity and the keplerian term $1/x$ dominates. Actually, this is an obvious consequence of the finite total mass of the model. Note, however, that the keplerian behaviour is attained only for large values of $x$ so that $v_c$ is quite slowly declining and, within the observational errors, may be also compatible with a flat rotation curve.

\begin{figure}
\resizebox{8.5cm}{!}{\includegraphics{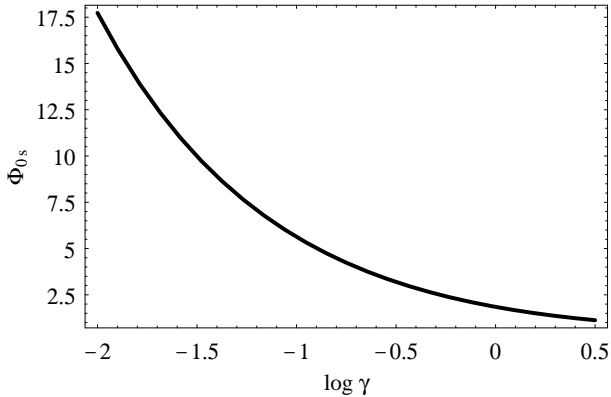}}
\caption{The scaled central potential $\Phi_{0s}$ as function of the logarithm of the slope parameter $\gamma$. We have defined  $\Phi_{0s} \equiv - \Phi(0)/(G M_{tot}/r_{-2})$.}
\label{fig: phiz}
\end{figure}

Let us now derive the gravitational potential for PoLLS models. To this aim, the Poisson equation\,:

\begin{displaymath}
\nabla^2 \Phi = 4 \pi G \rho(r) 
\end{displaymath}
has to be solved. For spherically symmetric system, the general solution is \cite{BT87}\,:

\begin{equation}
\Phi(r) = - \frac{G M(r)}{r} - 4 \pi G \int_{r}^{\infty}{\rho(r') r' dr'} \ .
\label{eq: genphi}
\end{equation}
Using Eq.(\ref{eq: mass}) for the mass profile and Eq.(\ref{eq: rho}) for the density law, we get\,:

\begin{equation}
\Phi(x) = - \Phi_{-2}(\rho_{-2}, r_{-2}, \gamma) \times \frac{{\cal{F}}(x; \gamma)}{{\cal{F}}(1; \gamma)}
\label{eq: phi}
\end{equation}
with\,:

\begin{equation}
\Phi_{-2} = \frac{G M_{tot}}{r_{-2}} \times {\cal{F}}(1; \gamma) \ ,
\label{eq: defphis}
\end{equation}

\begin{eqnarray}
{\cal{F}}(x; \gamma) & = & \frac{\Gamma(3/\gamma) - \Gamma(3/\gamma, 2 x^{\gamma}/\gamma)}{x \Gamma(3/\gamma)} + \nonumber \\
~ & + & \left ( \frac{2}{\gamma} \right )^{1/\gamma} \frac{\Gamma(2/\gamma) + \Gamma(2/\gamma, 2 x^{\gamma}/\gamma)}{\Gamma(3/\gamma)} \ .
\label{eq: defeffe}
\end{eqnarray}
Being the total mass finite, Eq.(\ref{eq: genphi}) easily shows that the gravitational potential vanishes at infinity. However, a caveat is in order here. If we consider the limit for $x \rightarrow \infty$ of Eq.(\ref{eq: phi}), we find $\Phi(\infty) = G M_{tot}/r_{-2} \times \Gamma(2/\gamma)/\Gamma(3/\gamma)$. Since the potential is defined by Eq.(\ref{eq: genphi}) up to an arbitrary additive constant, we could add this value to Eq.(\ref{eq: phi}) in order to have $\Phi(x)$ vanishing at infinity as it is indeed the case. At the opposite end, having the model a finite force at the centre, the gravitational potential does not diverge in the origin. By applying the limit for $x \rightarrow 0$ in Eq.(\ref{eq: phi}), we get\,:

\begin{equation}
\Phi(0) = - \frac{G M_{tot}}{r_{-2}} \left ( \frac{2}{\gamma} \right )^{1/\gamma} \frac{\Gamma(2/\gamma)}{\Gamma(3/\gamma)} 
\label{eq: phizero}
\end{equation}
which is plotted in Fig.\,\ref{fig: phiz}. The central potential turns out to be an increasing\footnote{Note that Fig.\,\ref{fig: phiz} report $- \Phi(0)$ so that a monotonically decreasing curve means an increasing central potential.} function of the slope parameter $\gamma$. As a consequence, the depth of the potential well is greater for larger values of $\gamma$ and, for these models, particles are more attracted toward the centre. Therefore, for a given $x$, the gravitational force and hence the number of stars within a distance $x$ from the centre is larger for larger values of $\gamma$ thus explaining why $M(x)/M_{tot}$ is an increasing function of the slope parameter. 

\begin{figure}
\resizebox{8.5cm}{!}{\includegraphics{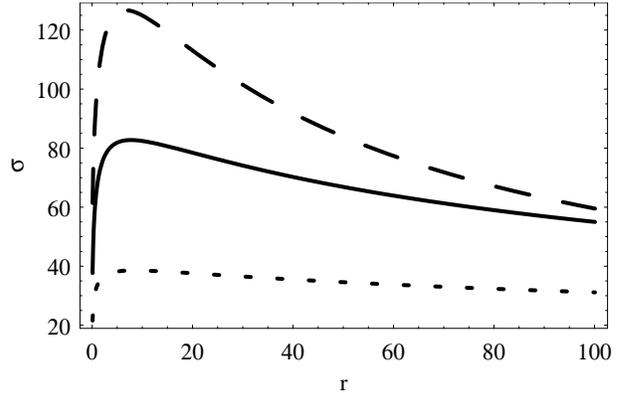}}
\caption{The velocity dispersion $\sigma$ (in km/s) as function of the radius $r$ (in kpc) for PoLLS models with three values of $\gamma$, namely $\gamma = 0.085$ (short dashed), $\gamma = 0.17$ (N04 model, solid) and $\gamma = 0.34$ (long dashed). The fiducial values $M_{tot} = 5 \times 10^{11} \ {\rm M_{\odot}}$ and $r_{-2} = 10 \ {\rm kpc}$ have been used.}
\label{fig: sigfig}
\end{figure}

Another dynamically interesting quantity that may be straightforwardly evaluated is the velocity dispersion. Assuming isotropy in the velocity space, this is given by \cite{BT87}\,:

\begin{displaymath}
\sigma^2(r) = \frac{1}{\rho(r)} \int_{r}^{\infty}{\rho(r') \frac{G M(r')}{r'^2} dr'} \ .
\end{displaymath}
Inserting Eqs.(\ref{eq: rho}) and (\ref{eq: mass}) into the above general expression, we get\,:

\begin{equation}
\sigma^2(x) = \frac{G M_{tot}}{r_{-2}} \times 
\frac{\exp{(- 2/\gamma)}}{\Gamma(3/\gamma)} \times I_{\sigma}(x; \gamma)
\label{eq: sigma}
\end{equation}
with 

\begin{equation}
I_{\sigma} \equiv \int_{x}^{\infty}{\left [ \Gamma(3/\gamma) - \Gamma(3/\gamma, 2 \xi^{\gamma}/\gamma) \right ] 
{\rm e}^{-(2/\gamma) \xi^{\gamma}} \xi^{-2} d\xi} \ .
\label{eq: defisig}
\end{equation}
This quantity may not be evaluated analytically, but it is straightforward to estimate numerically\footnote{In Appendix B we give analytical expressions that approximate very well some of the dynamical quantities we compute in this paper for the special case of the N04 model.} for fixed values of the slope parameter $\gamma$. Fig.\,\ref{fig: sigfig} shows the velocity dispersion for three values of $\gamma$ using $M_{tot} = 5 \times 10^{11} \ {\rm M_{\odot}}$ and $r_{-2} = 10 \ {\rm kpc}$. The result may be simply scaled to other values of $(M_{tot}, r_{-2})$ by noting that, because of Eq.(\ref{eq: sigma}), $\sigma \propto \sqrt{M_{tot}/r_{-2}}$. As yet observed for the rotational velocity, for a given $x$, $\sigma$ is higher for higher values of $\gamma$. In both cases, we have thus a degeneracy between the total mass and the slope parameter since both these quantities increase $v_c$ and $\sigma$. However, the degeneracy may be broken by determining where the rotation curve peaks since this only depends on $\gamma$.  As a further remark, we note that $\sigma$ decreases in the outer regions of the system thus mimicking well what has been recently observed in intermediate luminosity elliptical galaxies \cite{Nicola}. 

\begin{figure}
\resizebox{8.5cm}{!}{\includegraphics{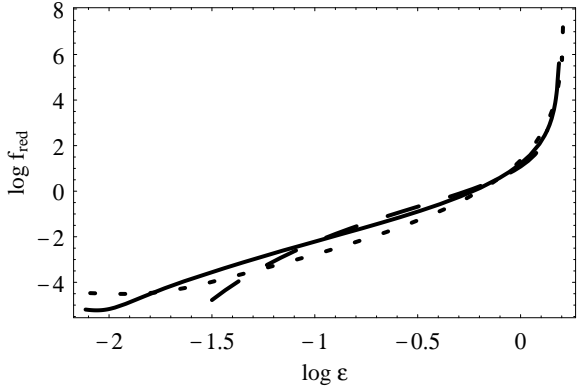}}
\caption{The logarithm of the reduced DF $f_{red}$ as function of the logarithm of the dimensionless binding energy $\hat{{\cal{E}}}$ for PoLLS models with three values of $\gamma$, namely $\gamma = 0.085$ (short dashed), $\gamma = 0.17$ (N04 model, solid) and $\gamma = 0.34$ (long dashed).}
\label{fig: dfgamma}
\end{figure}

\section{The distribution function}

All the kinematical information on a collisionless system may be derived by the phase space distribution function (hereafter DF) $f({\bf r}, {\bf v})$ that, for a non rotating spherically symmetric isotropic system, depends only on the binding energy for unit mass. Following Binney \& Tremaine (1987), we define\,:

\begin{displaymath}
\Psi(r) = -\Phi(r) + \Phi_0 \ \ , \ \ {\cal{E}}(r) = -\Psi(r) + \frac{1}{2} v^2(r)
\end{displaymath}
and set $\Phi_0 = 0$ so that the DF vanishes for ${\cal{E}} < 0$. The DF may then be obtained by means of the Eddington formula \cite{BT87}\,:

\begin{displaymath}
f({\cal{E}}) = \frac{1}{\sqrt{8} \pi^2} \frac{d}{d{\cal{E}}} 
\int_{0}^{{\cal{E}}}{\frac{d\rho}{d\Psi} \frac{d\Psi}{\sqrt{{\cal{E}} - \Psi}}} \ .
\end{displaymath}
In principle, one should express $\rho$ as function of $\Psi$ by first inverting the relation $\Psi = \Psi(x)$. Unfortunately, Eq.(\ref{eq: phi}) may not be inverted analytically so that we have to resort to a different approach changing variable from $\Psi$ to $x$ in the above Eddington formula. Using Eq.(\ref{eq: rho}) and (\ref{eq: phi}), we rewrite the Eddington formula as\,:

\begin{equation}
f(\hat{{\cal{E}}}) = \frac{2 \rho_{-2}}{\pi^2 \Phi_ {-2}\sqrt{8 \Phi_{-2}}} \frac{d}{d{\hat{\cal{E}}}}
\int_{x(\hat{{\cal{E}}})}^{\infty}{\frac{\xi^{\gamma - 1} \exp{\left [ - (2/\gamma) (\xi^{\gamma} - 1) \right ]}}
{\sqrt{\hat{{\cal{E}}} - \hat{\Psi}(\xi)}} d\xi}
\label{eq: eddfor}
\end{equation}
having introduced the dimensionless quantities\,:

\begin{displaymath}
\hat{{\cal{E}}} = {\cal{E}}/\Phi_{-2} \ \ , \ \ \hat{\Psi} = \Psi/\Phi_{-2} \ .
\end{displaymath}
In Eq.(\ref{eq: eddfor}) the lower limit of integration is implicitly defined by the equation\,:

\begin{equation}
\hat{\Psi}[x(\hat{{\cal{E}}})] = \hat{{\cal{E}}} 
\label{eq: solxi}
\end{equation}
that is easy to solve numerically. The dimensionless energy $\hat{{\cal{E}}}$ ranges between $|\Phi(\infty)/\Phi_{-2}|/$ and $|\Phi(0)/\Phi_{-2}|$ that, in our case, reduces to $0 \le \hat{{\cal{E}}} \le |\Phi(0)/\Phi_{-2}|$ with $\Phi(0)$ given by Eq.(\ref{eq: phizero}). For values of $\hat{{\cal{E}}}$ in this range, we may numerically determine the DF of the PoLLS models by first setting the value of the slope parameter $\gamma$, then solving Eq.(\ref{eq: solxi}) with respect to $x(\hat{{\cal{E}}})$ and finally using Eq.(\ref{eq: eddfor}). A caveat is in order here. For $\hat{{\cal{E}}}$ running in its range, $x$ changes between $0$ and $\infty$. Actually, when numerically solving Eq.(\ref{eq: solxi}), there are technical problems approaching the lower and upper ends of the range for the dimensionless binding energy. As a consequence, we compute the DF only over a limited range. The result is shown in Fig.\,\ref{fig: dfgamma} where we plot the logarithm of the reduced DF defined as\,:

\begin{displaymath}
f_{red}(\hat{{\cal{E}}}) = f(\hat{{\cal{E}}}) \times \left ( \frac{2 \rho_{-2}}{\pi^2 \Phi_{-2} \sqrt{8 \Phi_{-2}}} \right )^{-1} \ .
\end{displaymath}
It is worth noting that the reduced DF is well approximated by a single power law for a large range of values of the binding energy $\hat{{\cal{E}}}$. This is apparent from Fig.\,\ref{fig: dfgamma} where the logarithm of the reduced DF turns out to be a linear function of $\log{\hat{{\cal{E}}}}$ for $\log{\hat{{\cal{E}}}}$. However, both the exponent of the power law (i.e., the slope of the linear part of the plot) and the range over which this approximation works sufficiently well depend in a complicated way on the parameter $\gamma$. In particular, the range over which the reduced DF is a power law reduces with increasing $\gamma$. As a general remark, we note that the shape of the logarithm of the reduced DF is the same for different values of $\gamma$ with a linear part followed by a steep increase. In particular, for a fixed value of $\hat{{\cal{E}}}$, the higher is $\gamma$, the larger is the reduced DF. It is important to stress, however, that the power law approximation does not hold at the low and high energy limits. Actually, for $\hat{{\cal{E}}}$ approaching $|\Phi(0)/\Phi_{-2}|$, the DF diverges (as shown by the steep increase in Fig.\,\ref{fig: dfgamma}) and the divergence is strongest for smaller $\gamma$. On the other hand, $f_{red}$ drops to zero much steeper than a power law for $\hat{{\cal{E}}} \rightarrow 0$ (not shown in the plot), but, unfortunately, we have been unable to find an approximated formula relating the steepness of the decrease with the slope parameter $\gamma$.

\begin{figure}
\resizebox{8.5cm}{!}{\includegraphics{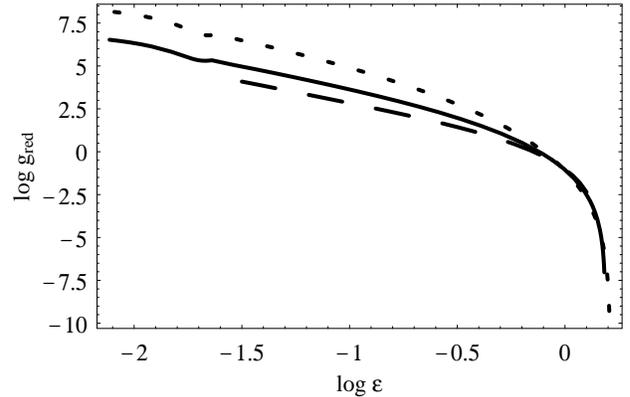}}
\caption{The logarithm of the reduced density of states $g_{red}$ as function of the logarithm of the dimensionless binding energy $\hat{{\cal{E}}}$ for PoLLS models with three values of $\gamma$, namely $\gamma = 0.085$ (short dashed), $\gamma = 0.17$ (N04 model, solid) and $\gamma = 0.34$ (long dashed).}
\label{fig: dngamma}
\end{figure}

The DF is a useful quantity,  but does not tell anything about the number of stars per unit binding energy. The quantity that gives this information is the differential energy distribution ${\cal{N}}({\cal{E}})$ that, for an isotropic and spherically symmetric system, may be written as ${\cal{N}}({\cal{E}}) = f({\cal{E}}) g({\cal{E}})$. Since we have yet determined the DF, computing the differential energy distribution reduces to the evaluation of $g({\cal{E}})$, that is the density of states per unit binding energy given by \cite{BT87}\,:

\begin{displaymath}
g({\cal{E}}) = 16 \sqrt{2} \pi^2 \int_{{\cal{E}}}^{\infty}{\left | r^2 \frac{dr}{d\Psi} \right | \sqrt{\Psi - {\cal{E}}} \ d\Psi} \ .
\end{displaymath}
Since $\Psi$ is a monotonic decreasing function of $r$, $dr/d\Psi < 0$ everywhere and we can rewrite the above relation as\,:

\begin{equation}
g(\hat{{\cal{E}}}) = 16 \sqrt{2 \Phi_{-2}} \pi^2 r_{-2}^3 \int_{0}^{x(\hat{{\cal{E}}})}{\sqrt{\Psi - {\cal{E}}} \ \xi^2 d\xi} \ 
\label{eq: dens}
\end{equation}
with $x(\hat{{\cal{E}}})$ the same as in Eq.(\ref{eq: solxi}). It is convenient to define a dimensionless reduced density of states as\,:

\begin{figure}
\resizebox{8.5cm}{!}{\includegraphics{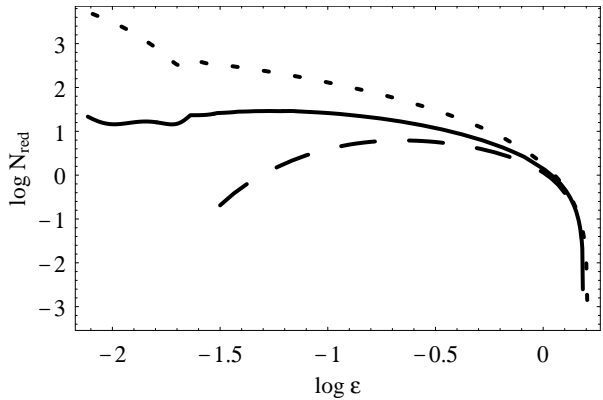}}
\caption{The logarithm of the reduced differential energy distribution $N_{red}$ as function of the logarithm of the dimensionless binding energy $\hat{{\cal{E}}}$ for PoLLS models with $\gamma = 0.085$ (short dashed), $\gamma = 0.17$ (N04 model, solid) and $\gamma = 0.34$ (long dashed).}
\label{fig: diffen}
\end{figure}

\begin{displaymath}
g_{red}(\hat{{\cal{E}}}) \equiv \frac{g(\hat{{\cal{E}}})}{16 \sqrt{2 \Phi_{-2}} \pi^2 r_{-2}^3} \ .
\end{displaymath}
As is shown in Fig.\,\ref{fig: dngamma}, the density of states may be approximated as a power law over a large range of $\hat{{\cal{E}}}$ as yet observed for the reduced DF, but now the slope is negative coherently with what is expected. The dependence on $\gamma$ is similar to that of the reduced DF with the only difference that the reduced density of states is a decreasing rather than an increasing function of the binding energy.

Finally, we show in Fig.\,\ref{fig: diffen} the differential energy distribution by plotting the logarithm of the reduced quantity\,:

\begin{displaymath}
{\cal{N}}_{red}(\hat{{\cal{E}}}) \equiv {\cal{N}}(\hat{{\cal{E}}}) \times \left ( \frac{ \Phi_{-2}}{16 \rho_{-2} r_{-2}^3} \right )^{-1} \ .
\end{displaymath}
${\cal{N}}$ turns out to be a decreasing function of the binding energy well approximated by a single power law (i.e. a linear fit in logarithmic space) over almost the full range of ${\cal{E}}$. The abrupt cutoff at high binding energies may be qualitatively explained by noting that high values of ${\cal{E}}$ are reached in the centre of the mass distribution where the density takes a finite value. As a result, the differential energy distribution may not go down forever, but has to be truncated. At the other extreme, since the mass density formally vanishes only at infinity, it is always possible to find stars at larger and larger distances with lower and lower binding energies so that ${\cal{N}}(\hat{{\cal{E}}})$ may diverge for ${\cal{E}}$ going to zero. This behaviour is apparently opposed to what is shown in Fig.\,\ref{fig: diffen} where a divergence appears only for the model with $\gamma = 0.085$, while the ${\cal{N}}_{red}$ is approximately constant for $\gamma = 0.17$ and decreasing for $\gamma = 0.34$. However, we have checked that these unexpected behaviours are the result of numerical problems originating for low values of the binding energy.

\section{Observed properties}

The density law in Eq.(\ref{eq: rho}) is a generalization of the N04 model that has been proposed to describe the dark haloes coming out from high resolution numerical simulations. Actually, there is no a priori reason against using PoLLS models to describe also the luminous components of galaxies (in particular, ellipticals and bulges of spiral galaxies). It is thus interesting to ask what is the surface mass density of these models since this may compared with the surface brightness. Moreover, this quantity directly enters the study of the lensing properties of a given model. It is thus worth computing the surface density of PoLLS models starting from the definition\,:

\begin{displaymath}
\Sigma(R) = \int_{0}^{R}{\frac{\rho(r) r dr}{\sqrt{r^2 - R^2}}}
\end{displaymath}
with $R$ the distance from the centre in the equatorial plane of the system. By inserting Eq.(\ref{eq: rho}) into the above relation, we get\,:

\begin{equation}
\Sigma(y) = \Sigma_{-2} \times \frac{{\cal{S}}(y, \gamma)}{{\cal{S}}(1, \gamma)}
\label{eq: sigmasd}
\end{equation}
with $y \equiv R/r_{-2}$ and we have defined\,:

\begin{equation}
{\cal{S}}(y, \gamma) \equiv \int_{0}^{y}{\frac{\exp{-(2/\gamma)(x^{\gamma - 1})} x dx}{\sqrt{x^2 - y^2}}} \ ,
\label{eq: defesse}
\end{equation} 

\begin{equation}
\Sigma_{-2} \equiv 2 \rho_{-2} r_{-2} {\cal{S}}(1, \gamma) \ .
\label{eq: sigm2}
\end{equation}
${\cal{S}}(y, \gamma)$ may not be evaluated analytically, but it is straightforward to compute numerically for fixed value of the slope parameter $\gamma$. The result is shown in Fig.\,\ref{fig: surf} where we use a logarithmic scale to make easier the visualization. It is worth noting that the slope parameter $\gamma$ plays a key role in determining the behaviour of $\Sigma$ in the inner regions, while has a still significant but less dramatic effect for large values of $y$. In particular, we find that the higher is $\gamma$, the lower is the surface density for a given $x$. This is particularly evident in the very inner regions as it is seen comparing the flat part of the curves in Fig.\,\ref{fig: surf}. It is worth stressing that the surface density may not mimic neither the usual $r^{1/4}$ law nor its generalization $r^{1/n}$. This is not a serious drawback of PoLLS models, however, since they were introduced to describe dark haloes rather than luminous components. It is nonetheless interesting to compute also the integrated surface density that is given as\,:

\begin{figure}
\resizebox{8.5cm}{!}{\includegraphics{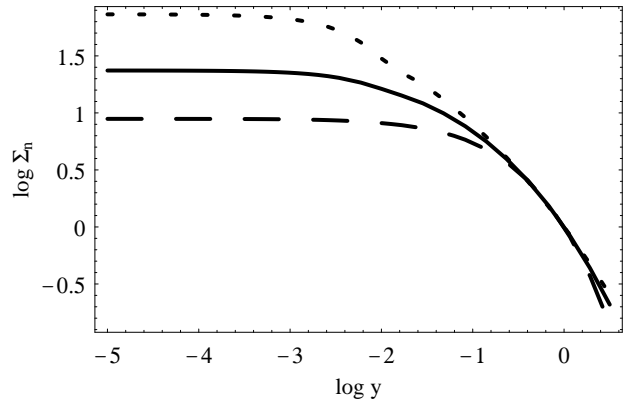}}
\caption{The logarithm of the scaled surface density $\Sigma_n \equiv \Sigma/\Sigma_{-2}$ as function of the logarithm of the scaled radius $y$ for PoLLS models with three values of $\gamma$, namely $\gamma = 0.085$ (short dashed), $\gamma = 0.17$ (N04 model, solid) and $\gamma = 0.34$ (long dashed).}
\label{fig: surf}
\end{figure}

\begin{equation}
S(y) = 2 \pi \int_{0}^{R}{\Sigma(R') dR'} =  
2 \pi \Sigma_{-2} r_{-2}^2 \int_{0}^{y}{\frac{{\cal{S}}(y', \gamma)}{{\cal{S}}(1, \gamma)} \ dy'}
\label{eq: sigmaint}
\end{equation}
that is shown in Fig.\,\ref{fig: surfint}. The dependence on $\gamma$ is further weakened by the integration process with the result that $S(y)$ is almost independent on the slope parameter up to $y \sim 2$. Note, however, that both $\Sigma_n$ and $S_n = S/(2 \pi \Sigma_{-2} r_{-2}^2)$ are increasing function of $\gamma$ as it is expected since models with smaller values of $\gamma$ are more concentrated, i.e. the projected mass within a given distance from the centre is larger.

In Sect.\,3, we have evaluated the velocity dispersion, but indeed it is not this quantity that is measured. Actually, from the observations it is possible to estimate the velocity dispersion projected along the line of sight and weighted with the surface luminosity density. This is given by\,:

\begin{equation}
\sigma_{los}^2 = \frac{2}{I(R)} \int_{R}^{\infty}{\rho_{stars}(r) \frac{G M_{T}(r) \sqrt{r^2 - R^2}}{r^2} dr}
\label{eq: sigmalos}
\end{equation}
where $I(R)$ is the total surface luminosity density, $\rho_{stars}$ the mass density of the stellar component and $M_{T}$ the total (stellar plus dark matter) mass profile. To be more realistic, we average over a finite radial bin $(R_1, R_2)$ as follows (see, e.g., the appendix in Dalal \& Keeton 2003)\,:

\begin{equation}
\sigma_{bin}^2 = \frac{\int_{R_1}^{R_2}{\rho_{stars}(r) G M_T(r) \left [ F(R_2/r) - F(R_1/r) \right ] dr}}
{2 \int_{R_1}^{R_2}{\rho_{stars}(r) \left [ A(R_2/r) - A(R_1/r) \right ] r dr}} 
\label{eq: sigmabin}
\end{equation}
with\,:

\begin{figure}
\resizebox{8.5cm}{!}{\includegraphics{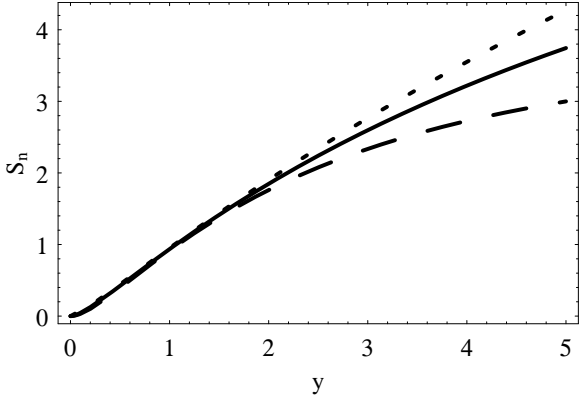}}
\caption{The normalized integrated surface density $S_n = S/(2 \pi \Sigma_{-2} r_{-2}^2)$ as function of the scaled radius $y$ for PoLLS models with three values of $\gamma$, namely $\gamma = 0.085$ (short dashed), $\gamma = 0.17$ (N04 model, solid) and $\gamma = 0.34$ (long dashed).}
\label{fig: surfint}
\end{figure}

\begin{equation}
A(x) = \left \{
\begin{array}{ll}
\arcsin(x) & x < 1 \\
~ & ~ \\
\pi/2 & x > 1 \\
\end{array}
\right . \ ,
\label{eq: defA}
\end{equation}

\begin{equation}
F(x) = \left \{
\begin{array}{ll}
x \sqrt{1 - x^2} + \arcsin(x) - \pi/2 & x < 1 \\
~ & ~ \\
0 & x > 1 \\
\end{array}
\right . \ .
\label{eq: defF}
\end{equation}
A caveat is in order here. In principle, $I(R)$ is not equal to $\Sigma(R)$. Actually, we have to distinguish among three possibilities. First, let us suppose that we use a PoLLS model for the dark halo of a galaxy in which case $\Sigma(R)$ does not enter in $I(R)$, while the mass profile of the model still enters in determining the velocity dispersion through $M_T$ in Eqs.(\ref{eq: sigmalos}) and (\ref{eq: sigmabin}). On the other hand, a PoLLS model may also be used to fit the surface brightness profile of an elliptical galaxy so that we need both $\Sigma$, $\rho$ and $M$ to evaluate $\sigma_{los}$ and $\sigma_{bin}$ and also choose a model for the dark halo. Finally, if the galaxy is a spiral one, we have a three component system thus complicating a lot describing the whole system and computing the velocity dispersion. In order to not introduce degeneracy among the model parameters and to better investigate the properties of the PoLLS models, we restrict our attention to elliptical galaxies and assume that it is possible to use a PoLLS model to fit the surface brightness. Moreover, we arbitrarily set $R_2 = 1.1 R_1$ which implies that the larger is the distance from the centre, the larger is the bin width thus allowing to reproduce qualitatively the need of averaging over larger bins in the outer regions to ameliorate the signal to noise ratio. 

Under these assumptions, $\rho_{stars}$ is given by Eq.(\ref{eq: rho}) and $M_T(r)$ by Eq.(\ref{eq: mass}) so that Eq.(\ref{eq: sigmabin}) becomes\,:

\begin{figure}
\resizebox{8.5cm}{!}{\includegraphics{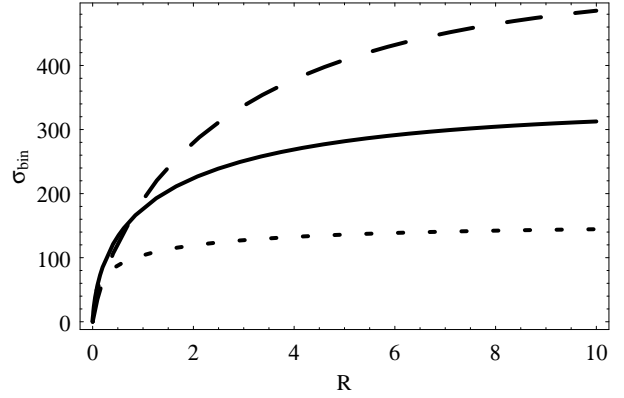}}
\caption{The projected luminosity weighted velocity dispersion averaged over a radial bin $\sigma_{bin}$ (in km/s) as function of the distance $R$ (in kpc) for PoLLS models with three values of $\gamma$, namely $\gamma = 0.085$ (short dashed), $\gamma = 0.17$ (N04 model, solid) and $\gamma = 0.34$ (long dashed). The fiducial values $M_{tot} = 5 \times 10^{11} \ {\rm M_{\odot}}$ and $r_{-2} = 10 \ {\rm kpc}$ have been used.}
\label{fig: sbgamma}
\end{figure}

\begin{equation}
\sigma_{bin}^2(y_1) = \frac{G M_{tot}}{2 r_{-2} \Gamma(3/\gamma)} \times I_{bin}(y_1) 
\label{eq: sigmabinpolls}
\end{equation}
with\,:

\begin{equation}
I_{bin}(y_1) = {\cal{N}}_{bin}(y_1)/{\cal{D}}_{bin}(y_1) \ ,
\label{eq: ibin}
\end{equation}

\begin{displaymath}
{\cal{N}}_{bin}(y_1) = \int_{y_1}^{1.1 y_1}{\left [ \Gamma(3/\gamma) - \Gamma(3/\gamma, 2 x^{\gamma}/\gamma) \right ] \ \times}
\end{displaymath}

\begin{displaymath}
\ \ \ \ \ \ \ \ \ \ \ \ \ \ \ \ \ \ \ \ \ \ \ \ \ 
\exp{\left [ - (2/\gamma)(x^{\gamma} - 1) \right ]} \ \times
\end{displaymath}

\begin{equation}  
\ \ \ \ \ \ \ \ \ \ \ \ \ \ \ \ \ \ \ \ \ \ \ \ \ 
\left [ F(1.1 y_1/x) - F(y_1/x) \right ] dx  \ ,
\label{eq: nbin}
\end{equation}

\begin{displaymath}
{\cal{D}}_{bin}(y_1) =
\int_{y_1}^{1.1 y_1}{x \exp{\left [ - (2/\gamma)(x^{\gamma} - 1) \right ]} \ \times} 
\end{displaymath}

\begin{equation}  
\ \ \ \ \ \ \ \ \ \ \ \ \ \ \ \ \ \ \ \ \ \ \ \ \ 
\left [ A(1.1 y_1/x) - A(y_1/x) \right ] dx \ . 
\label{eq: dbin}
\end{equation}
Not surprisingly, we are unable to give an analytical expression for $I_{bin}$, but it is not difficult to evaluate it numerically for fixed value of the slope parameter $\gamma$. To get some interesting examples, we use the same fiducial model adopted for computing $\sigma$ in Fig.\,\ref{fig: sigfig} obtaining the plot in Fig.\,\ref{fig: sbgamma}. Noting that, by virtue of Eq.(\ref{eq: sigmabinpolls}), $\sigma_{bin}$ is proportional to $\sqrt{M_{tot}/r_{-2}}$, it is immediate to scale the results for the fiducial model to other values of these two parameters. It is worth noting that there is not a clear trend of $\sigma_{bin}$ with $\gamma$ in the very inner regions ($y_1 < 0.05$ not visible in the plot) as a result of the complicated interplay between the density and the mass profiles due to the fact that $M(r)$ enters only in ${\cal{N}}_{bin}(y_1)$. However, for larger values of $y_1$, we recover the same behaviour yet observed for the velocity dispersion $\sigma$ shown in Fig.\,\ref{fig: sigfig} with $\sigma_{bin}$ increasing with $\gamma$. Fig.\,\ref{fig: sbgamma} does not extend further than $y = 0.1$. While there are no technical difficulties in computing $\sigma_{bin}$ for $y \ge 0.1$ thus obtaining a declining projected velocity dispersion, Eqs.(\ref{eq: sigmabinpolls})\,-\,(\ref{eq: dbin}) rigorously do not hold anymore in this regime since they have been obtained under the hypotheses that there is no dark matter (or that its contribution may be neglected) and a PoLLS model may describe the luminous component. Both these hypotheses (in particular, the former one) introduce systematics that quickly increase approaching the outer regions of the galaxy so that we have preferred to only evaluate $\sigma_{bin}$ in the very interiors to reduce these errors.

\section{Anisotropic models}

The models in the family we have described have an isotropic dynamical structure, i.e. we have assumed that the velocity dispersion is the same along the three axes of the velocity ellipsoid. Actually, this hypothesis is hardly verified in real systems so that it is important to extend our models to take into account a possible anisotropy in the orbital structure. The radial velocity dispersion $\sigma_r$ is determined by solving the Jeans equation that now writes\,:

\begin{equation}
\frac{d(\rho \sigma_r^2)}{dr} + 2 \frac{\beta_{\sigma}(r)}{r} \rho(r) \sigma_r^2(r) = - \rho(r) \frac{G M(r)}{r^2} 
\label{eq: jeans}
\end{equation} 
with\

\begin{equation}
\beta_{\sigma}(r) \equiv 1 - \frac{\sigma_r^2}{\sigma_t^2} 
\label{eq: defbeta}
\end{equation} 
where $\sigma_t$ is the tangential velocity dispersion. In the isotropic case, $\sigma_t = \sigma_r = \sigma$ and Eq.(\ref{eq: sigma}) holds. In the general case, the solution of Eq.(\ref{eq: jeans}) is \cite{BT87,ML04b}\,: 

\begin{equation}
\sigma_r^2(r) = \frac{1}{\eta(r) \rho(r)} \int_{r}^{\infty}{\eta(r') \rho(r') \frac{G M(r')}{r'^2} dr'}
\label{eq: anisig}
\end{equation}
with $\eta(r)$ related to the anisotropy parameter $\beta_{\sigma}$ as\,:

\begin{equation}
\frac{d\log{\eta}}{d\log{r}} = 2 \beta_{\sigma}(r) \ .
\label{eq: etasol}
\end{equation}
The estimate of the anisotropy parameter $\beta_{\sigma}$ is quite difficult so that different choices are possible. A widely used parameterization is the Osipkov\,-\,Merrit one \cite{O79,M85}\,:

\begin{equation}
\beta_{\sigma}(x) = \frac{x^2}{x^2 + x_{OM}^2} 
\label{eq: ombeta}
\end{equation}
that is characterized by the anisotropy radius $r_{OM} = r_s x_{OM}$. Solving Eq.(\ref{eq: etasol}) with $\beta_{\sigma}$ given by Eq.(\ref{eq: ombeta}) and inserting the result and Eqs.(\ref{eq: rho}) and(\ref{eq: mass}) into the general expression (\ref{eq: anisig}), we get\,:

\begin{displaymath}
\sigma_r^2 = \frac{G M_{tot}}{r_{-2}} \times \left \{ \Gamma(3/\gamma) (x^2 + x_{OM}^2) \exp{\left [ - (2/\gamma) (x^{\gamma} - 1) \right ]} \right \}^{-1}
\end{displaymath}

\begin{equation}
\ \ \ \ \ \ \ \ \ \ \ \ \ \ \ \ \ \ \ \ \   
\times \ I_{OM}(x; \gamma, x_{OM})  
\label{eq: sigmaom}
\end{equation}
with\,:

\begin{displaymath}
I_{OM} = \int_{x}^{\infty}{\left [ \Gamma(3/\gamma) - \Gamma(3/\gamma, 2 \xi^{\gamma}/\gamma) \right ]} \ \times
\end{displaymath}

\begin{equation} 
\ \ \ \ \ \ \ \ \ \ \ \ \ \ \ \ 
\exp{\left [ - (2/\gamma) (\xi^{\gamma} - 1) \right ]} \ \frac{\xi^2 + x_{OM}^2}{\xi^2} \ d\xi \ .
\label{eq: defiom}
\end{equation}
Even if widely used, it has been recently shown that the Osipkov\,-\,Merritt parameterization does not fit well the anisotropy parameter measured in cosmological N\,-\,body simulations \cite{Dia99,Col00,RTM,Die04}.  Indeed, Mamon \& Lokas (2004a,b) have shown that the following parameterization works better\,:

\begin{equation}
\beta_{\sigma}(x) = \frac{x}{2 (x + x_{ML})}
\label{eq: lmbeta}
\end{equation} 
with $x_{ML} = r_{ML}/r_s$ and $r_{ML}$ a characteristic anisotropy radius. With the Mamon \& Lokas parameterization of the anisotropy parameter, the radial velocity dispersion turns out to be\,:

\begin{figure}
\resizebox{8.5cm}{!}{\includegraphics{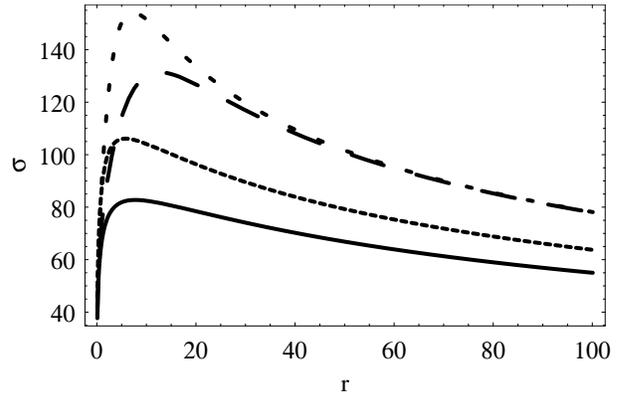}}
\caption{The radial velocity dispersion $\sigma$ (in km/s) of the N04 model as function of the radius $r$ (in kpc) for the isotropic model (solid), the Osipkov\,-\,Merritt models with $x_{OM} = 0.5$ (short dashed) and $x_{OM} = 1$ (long dashed) and the Mamon \& Lokas model with $x_{ML} = 0.18$ (continuous dashed). The fiducial values $M_{tot} = 5 \times 10^{11} \ {\rm M_{\odot}}$ and $r_{-2} = 10 \ {\rm kpc}$ have been used.}
\label{fig: anisigma}
\end{figure}

\begin{displaymath}
\sigma_r^2(x) = \frac{G M_{tot}}{r_{-2}} \times \left \{ \Gamma(3/\gamma) (x + x_{ML}) \exp{\left [ - (2/\gamma) (x^{\gamma} - 1) \right ]} \right \}^{-1}
\end{displaymath}

\begin{equation}
\ \ \ \ \ \ \ \ \ \ \ \ \ \ \ \ \ \ \ \ \   
\times \ I_{ML}(x; \gamma, x_{ML})  
\label{eq: sigmaml}
\end{equation}
with\,:

\begin{displaymath}
I_{ML} = \int_{x}^{\infty}{\left [ \Gamma(3/\gamma) - \Gamma(3/\gamma, 2 \xi^{\gamma}/\gamma) \right ]} \ \times
\end{displaymath}

\begin{equation} 
\ \ \ \ \ \ \ \ \ \ \ \ \ \ \ \ 
\exp{\left [ - (2/\gamma) (\xi^{\gamma} - 1) \right ]} \ \frac{\xi + x_{ML}}{\xi^2} \ d\xi \ .
\label{eq: defiml}
\end{equation}
Actually, it is worth noting that some of the simulations considered by Mamon \& Lokas are tailored on cluster scale haloes \cite{RTM} or on the subhaloes \cite{Die04} so that it is at least premature to reject the Osipkov\,-\,Merritt model for galaxy scale systems. Hence, in Fig.\,\ref{fig: anisigma}, we report the radial velocity dispersion $\sigma_r$ for the N04 model for the isotropic model, the Osipkov\,-\,Merrit and the Mamon \& Lokas anisotropic parameterizations. As expected, anisotropic models predict an higher velocity dispersion with the Osipkov\,-\,Merritt scheme being more effective in raising $\sigma_r$. In particular, it turns out that the lower is $x_{OM}$, the higher is the the radial velocity dispersion that may be qualitatively explained by noting that the model becomes more and more isotropic as $x_{OM}$ increases. We have checked that these results do not depend on $\gamma$. Note that there is a degeneracy between mass and anisotropy so that it is difficult to choose among isotropic, Osipkov\,-\,Merritt and Mamon \& Lokas models on the basis of the radial velocity dispersion only. Moreover, the differences among these models may be partially washed out when considering the projected velocity dispersion since this enters also other parameters (related to the luminous components) thus leading to possible degeneracies that seriously complicate the interpretation of the data. 

Introducing an anisotropy in the velocity space also changes the DF of the model that now becomes a function not only of the binding energy ${\cal{E}}$, but also of the modulus $L$ of the angular momentum vector. The construction of general anisotropic DF $f({\cal{E}}, L)$ for spherically symmetric potential\,-\,density pairs is discussed in detail in Dejonghe (1986). However, here, we restrict our attention to the Osipkov\,-\,Merrit parameterization in which case the DF depends on ${\cal{E}}$ and $L$ only through the parameter\,:

\begin{figure}
\resizebox{8.5cm}{!}{\includegraphics{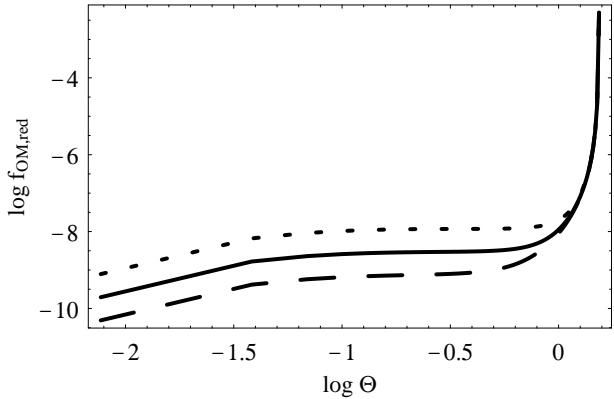}}
\caption{The logarithm of the reduced DF of the N04 model as function of $\log{\Theta}$ (with $\Theta = \hat{{\cal{Q}}}$) for three values of the dimensionless anisotropy radius $x_{OM}$, namely $x_{OM} = 0.5$ (short dashed), $x_{OM} = 1.0$ (solid) and $x_{OM} = 2.0$ (long dashed).} 
\label{fig: anidf}
\end{figure}

\begin{equation}
{\cal{Q}} \equiv {\cal{E}} - \frac{L^2}{2 r_{OM}^2} \ .
\label{eq: ompar}
\end{equation}
It is then possible to compute the DF using the following generalization of the Eddington formula \cite{D86,BT87}\,:

\begin{equation}
f({\cal{Q}}) = \frac{1}{\sqrt{8} \pi^2} \int_{0}^{{\cal{Q}}}{\frac{d^2 \rho_{{\cal{Q}}}}{d\Psi^2} \frac{d\Psi}{\sqrt{{\cal{Q}} - \Psi}}}
\label{eq: dfom}
\end{equation}
with

\begin{equation}
\rho_Q(r) = \left [ 1 + \left ( \frac{r}{r_{OM}} \right )^2 \right ] \rho(r) \ .
\label{eq: defrhoq}
\end{equation}
Defining the dimensionless parameter $\hat{{\cal{Q}}} = {\cal{Q}}/\Phi_s$ and changing variable, we get for PoLLS models\,:

\begin{equation}
f(\hat{{\cal{Q}}}) = - \frac{\rho_{-2}}{\sqrt{8 \Phi_{-2}} \pi^2 \Phi_{-2}} \times f_{OM,red}(\hat{{\cal{Q}}}) 
\label{eq: anidfpolls}
\end{equation}
where we have defined\,:

\begin{displaymath}
f_{OM,red} =
\int_{x_Q(\hat{{\cal{Q}}})}^{\infty}
{\left \{ \left ( \frac{d\hat{\Psi}}{dx} \right )^{-1} \frac{d^2 \hat{\rho}_Q}{dx^2} + 
\frac{d \hat{\rho}_Q}{dx} \frac{d}{dx} \left [ \left ( \frac{d \hat{\Psi}}{dx} \right )^{-1} \right ] \right \}} \ \times
\end{displaymath}

\begin{equation}
\ \ \ \ \ \ \ \ \ \ \ \ \ \ \ \ \ \ \ \ \ \ \ \ 
\frac{dx}{\sqrt{\hat{{\cal{Q}}} - \hat{\Psi}(x)}}
\label{eq: omdf}
\end{equation}
with $\hat{\rho}_Q = \rho_{Q}/\rho_s$ and $x_Q(\hat{{\cal{Q}}})$ obtained by solving Eq.(\ref{eq: solxi}) after replacing $\hat{{\cal{E}}}$ with $\hat{{\cal{Q}}}$. Not surprisingly, the reduced DF $f_{OM,red}$ may not be evaluated analytically, but it is straightforward to compute numerically. The result for the N04 model is shown in Fig.\,\ref{fig: anidf} for three arbitrarily chosen values of the anisotropy radius. 
Comparing with Fig.\,\ref{fig: dfgamma} for the isotropic case, we note that the reduced DF is no more approximated by a simple power law. The dependence on the anisotropy dimensionless radius $x_{OM}$ is quite strong for low values of $\hat{{\cal{Q}}}$ with the reduced DF being larger for smaller $x_{OM}$. On the other hand, there is almost no dependence at all on $x_{OM}$ for high values of $\hat{{\cal{Q}}}$ since, in this regime, it is $\hat{{\cal{E}}} >> L/2r_{OM}^2$ so that $\hat{{\cal{Q}}} \simeq \hat{{\cal{E}}}$ and we reduce to the isotropic case. Moreover, whatever is the value of $r_{OM}$, the DF is divergent at the upper end of $\hat{{\cal{Q}}}$ (as in the isotropic case with $\hat{{\cal{Q}}}$ replacing $\hat{{\cal{E}}}$) so that different values of the anisotropy radius have a negligible impact.

\section{Conclusions}

In an attempt to look for underlying similarities among the plethora of spherically symmetric galaxy models discussed in literature, we have shown that most of the cuspy models may be considered as belongings to a general family defined by Eq.(\ref{eq: genslope}) for the logarithmic slope $\alpha(r)$ of the radial profile. We have then restricted our attention to a subclass of models, dubbed PoLLS, whose logarithmic slope is a power law in $r$ since this is in agreement with the results of the most recent and detailed numerical simulations available up to now \cite{Nav04}.   

PoLLS models are physically meaningful only if the slope parameter $\gamma$ is positive. The mass density is non singular in the centre so that the models are not cuspy, while $\rho$ is an exponentially decreasing function of $r$ so that the total mass turns out to be finite. This is an interesting feature of the model that makes it different from, e.g., the popular NFW model (Navarro et al. 1996, 1997) that has to be truncated at the virial radius in order to define a finite total mass. The mass profile, the rotation curve and the gravitational potential of PoLLS models have been evaluated analytically and turn out to be expressed in terms of special functions that are quite easy to handle (e.g., in numerical codes). In particular, while being asymptotically keplerian, the rotation curve is only slowly declining so that PoLLS models may be good candidates for the dark halo of spiral galaxies. Moreover, the position of the peak of the rotation curve may be used to constrain the value of the slope parameter thus breaking the degeneracy between $\gamma$ and the total mass that have the same effect on both the rotation curve and the velocity dispersion.

Assuming isotropy in the velocity space, we have numerically evaluated the velocity dispersion that turns out to decline in the outer regions of the system thus offering the intriguing possibility that such models may reconcile the presence of a dark halo with the apparent lack of dark matter suggested by the declining velocity dispersion measured in intermediate luminosity elliptical galaxies \cite{Nicola}. Actually, it is not this quantity that has to be compared with the data, but the luminosity weighted line of sight velocity dispersion that we have computed assuming that the PoLLS model is the only galaxy component. This is, of course, a quite unrealistic assumption so that a detailed comparison will be performed in a forthcoming paper where we will also take care of the surface brightness to constrain the modelling of luminous components.

Being the models spherically symmetric and isotropic, the DF only depends on the binding energy ${\cal{E}}$ and may be estimated using the Eddington formula. It turns out that the DF may be approximated as a single power law over a quite large range for ${\cal{E}}$, but the width of this interval and the goodness of the approximation depend critically on the values of the slope parameter $\gamma$. The density of states has also been evaluated for completeness and shows some qualitative features similar to the DF, but with the opposite dependence on ${\cal{E}}$ being a decreasing rather than an increasing function of the binding energy. The product of the DF and the density of states gives the differential energy distribution ${\cal{N}}$ that turns out to be a decreasing function with an abrupt cutoff at the upper extreme of the range for ${\cal{E}}$ due to the finite central density.
 
The properties of the PoLLS models have all been investigated by assuming both spherical symmetry and isotropy in the velocity space. As a first step towards a more realistic description, we have also considered the effects of non vanishing anisotropy parameter $\beta_{\sigma}$ on the radial velocity dispersion exploring two different parameterization, namely the popular Osipkov\,-\,Merrit model \cite{O79,M85} and the one recently proposed by Mamon \& Lokas (2004a,b). In both cases, introducing an anisotropy in the velocity space leads to an increase of the radial velocity dispersion with respect to the corresponding isotropic models thus originating the well known mass\,-\,anisotropy degeneracy. Actually, a combined analysis of both the line of sight velocity dispersion and the rotation curve could help in breaking this degeneracy being $v_c$ insensitive to $\beta_{\sigma}$. 

We have also estimated the DF for Osipkov\,-\,Merrit anisotropic PoLLS models finding out that the power law approximation for the DF breaks down. As it is expected, the DF now depends also on the anisotropy radius $r_{OM}$, but this dependence cancels out when considering high values of ${\cal{Q}}$ since the higher is ${\cal{Q}}$, the more isotropic is the model and the nearer is the DF to its divergence at ${\cal{Q}}_{max}$.

PoLLS models are a first attempt towards a different approach to galaxy modelling. Rather than starting from an expression for the mass density, our initial guess was on the logarithmic slope of the radial profile. This quantity gives an immediate understanding of the asymptotic behaviours of the density profile. Moreover, it seems that $\alpha(r)$ is better constrained from numerical simulations. While our attention on PoLLS models was mainly motivated by the results of Navarro et al. (2004) that refer to dark haloes, modelling a galaxy also requires to consider its luminous components. These latter may be efficiently described by Zhao models (1996, 1997) that represent a generalization of both the popular Jaffe (1983) and Hernquist (1990) profiles. Kinematical and dynamical data may then be used to constrain the properties of the full galaxy modeled using a Zhao model for the luminous component and a PoLLS one for the dark halo. 

We would like to conclude with a general comment. As the processing speed of modern supercomputers evolves, introducing new (sophisticated or not) models to fit the results of increasingly detailed numerical simulations becomes a more and more attractive temptation. Actually, we believe that time is come to look for similarities rather than discrepancies among the models in order to find out what is the right answer to the questions posed by the structure of numerically simulated haloes. In our opinion, approaching the problem starting from the logarithmic slope may be a good road to successfully arrive at the final destination. 

\section*{acknowledgements}

We warmly thank an anonymous referee for drawing our attention to the $(\alpha, \beta, \gamma)$ models and for constructive comments that have helped us to significantly improve the paper. CT is financially funded by the P.R.I.N. ``Draco'' of the Italian Ministry of Education, University and Research.

\begin{table*}
\caption{Density profile (with $x = r/r_s$ and $r_s$ a typical scale radius) and values of the parameters $(\beta, a, b, c, d, \gamma)$ for some popular spherical galaxy models. In the last column, we also give some useful references.}
\begin{center}
\begin{tabular}{|c|c|c|c|c|c|c|c|c|} 
\hline
Model & $\rho$ & $\beta$ & $a$ & $b$ & $c$ & $d$ & $\gamma$ & Ref. \\
\hline
Isothermal & $x^{-2}$ & 2 & 0 & 0 & 0 & 0 & 0 & Binney \& Tremaine 1987 \\ 
Jaffe & $x^{-2} (1 + x)^{-2}$ & 2 & 2 & 0 & 1 & 0 & 0 & Jaffe 1983 \\
Hernquist & $x^{-1} (1 + x)^{-3}$ & 1 & 4 & 0 & 1 & 0 & 0 & Hernquist 1990 \\
Denhen & $x^{-\gamma_D} (1 + x)^{\gamma_D - 4}$ & $\gamma_D$ & $4/\gamma_D$ & 0 & 1 & 0 & 0 & Dehnen 1993 \\
NFW & $x^{-1} (1 + x)^{-2}$ & 1 & 3 & 0 & 1 & 0 & 0 & Navarro et al. 1996, 1997 \\
Generalized NFW & $x^{-\beta_{GNFW}} (1 + x)^{\beta_{GNFW} - 3}$ & $\beta_{GNFW}$ & $3/\beta_{GNFW}$ & 0 & 1 & 0 & 0 & Jiing \& Suto 2000 \\
Cuspy halo & $x^{-\gamma_{CH}} (1 + x^2)^{(\gamma_{CH} - n)/2}$ & $\gamma_{CH}$ & 0 & $n/\gamma_{CH}$ & 0 & 1 & 0 & Mu\~{n}oz et al. 2001 \\
N04 & see Eq.(\ref{eq: rho}) & 2 & 0 & 0 & 0 & 0 & 0.17 & Navarro et al. 2004 \\
RTM & $x^{-1} (1 + x)^{-3/2}$ & 1 & 5/2 & 0 & 1 & 0 & 0 & Rasia et al. 2004 \\ 
\hline
\end{tabular}
\end{center}
\end{table*}

\appendix

\section{Logarithmic slope for some known models}

Eq.(\ref{eq: genslope}) is a general expression for the logarithmic slope of many spherically symmetric models. To give a quantitative demonstration of this ansatz, we report in Table\,1 the values of the parameters $(\beta, a, b, c, d, \gamma)$ for some cuspy models discussed in literature. It is worth noting that for all the models it is $b = d = 0$ and $c = 1$ with only two remarkable exceptions. The first is the {\it cuspy halo model} \cite{MKK01} that is actually used in lensing applications, but has not been studied dynamically. The second case is the N04 model \cite{Nav04} that belongs to the class of PoLLS models we have discussed here. 

\section{Approximated formulae for the N04 model}

The basic properties (such as the mass profile, the rotation curve and the gravitational potential) of PoLLS models may be expressed analytically in terms of special functions, while this is not for some other interesting quantities as the velocity dispersion and the DF. These quantities may be generally expressed as the product of a scaling factor that only depends on $r_{-2}$ and $M_{tot}$ and a definite integral that can be evaluated numerically for fixed values of the slope parameter $\gamma$. To save computing time in applications, we give here analytical approximated formulae for the N04 model for which it is $\gamma = 0.17$. 

Let us first consider the isotropic velocity dispersion which may be factorized as 

\begin{displaymath}
\sigma^2(x) = \frac{G M_{tot}}{r_{-2}} \times \frac{\exp{(- 2/\gamma)}}{\Gamma(3/\gamma)} \times I_{\sigma}(x; \gamma)
\end{displaymath}
with $I_{\sigma}$, defined in Eq.(\ref{eq: defisig}), to be computed numerically. For the N04 model, this latter quantity may be approximated as\,:

\begin{equation}
\log{I_{\sigma}(x; \gamma = 0.17)} \simeq \sum_{n = 0}^{3}{a_n (\log{x})^n}
\label{eq: sigapprox}
\end{equation}
with\,:

\begin{displaymath}
(a_0, a_1, a_2, a_3) = (7.50404, -2.06006, -0.66146, -0.07771) \ .
\end{displaymath}
This approximation works quite well with the error $\Delta I_{\sigma}/I_{\sigma}^{true}$ (being $\Delta I_{\sigma} = I_{\sigma}^{true} - I_{\sigma}^{fit}$) which is lower than $\sim 0.5\%$ for $0 \le x \le 0.1$ and even smaller (lower than $\sim 0.08\%$) in the range (0.1, 10).

A fundamental quantity to study the dynamical properties of a galaxy model is the DF. The reduced DF of the isotropic N04 model may be approximated as\,:

\begin{equation}
\log{f_{red}}(\hat{{\cal{E}}}) \simeq 1.2052 (1 - \varepsilon)^{-0.338} \sum_{n = 0}^{n = 3}{e_n \varepsilon^n}
\label{eq: approxdf}
\end{equation}
with $\varepsilon \equiv \log{\hat{{\cal{E}}}}/\log{\hat{{\cal{E}}}_{\star}}$, $\hat{{\cal{E}}}_{\star} = 1.614$ and\,:
\begin{displaymath}
(e_0, e_1, e_2, e_3) = (1, 0.923, 0.0275, 0.00476) \ .
\end{displaymath}
The error is lower than $4\%$ on $f_{red}$ for $-2 \le \log{\hat{{\cal{E}}}} \le 0.15$ so that the approximation is indeed quite good. It is worth stressing, however, that Eq.(\ref{eq: approxdf}) must not be used outside the range quoted since leads to serious systematic errors. For instance, the approximated reduced DF does not diverge for $\hat{{\cal{E}}} \rightarrow \hat{{\cal{E}}}_{max}$ as it is for the the true reduced DF.

Finally, let us consider the case of the surface density $\Sigma$ that we may write as $\Sigma_s \times \Sigma_n$ with $\Sigma_n = {\cal{S}}(y; \gamma)/{\cal{S}}(1; \gamma)$ numerically evaluated from Eq.(\ref{eq: sigmasd}). It is convenient to use the logarithm of $\Sigma_n$ (that is immediately related to the surface brightness) rather $\Sigma_n$ itself when looking for an approximated analytical expression. For the N04 model, we have found out that\,:

\begin{equation}
\log{\Sigma_n(y; \gamma = 0.17)} \simeq \sum_{n = 0}^{4}{s_n (\log{y})^n}
\label{eq: logsigapprox}
\end{equation}
with\,:

\begin{displaymath}
(s_0, s_1, s_2, s_3) = (-0.002, -1.159, -0.369, -0.369, -0.0030) \ ,
\end{displaymath}
reproduces $\Sigma_n$ with an error lower than $\sim 2.5\%$ over the full range $y = R/r_{-2} \in (0, 0.1)$, being $\Delta \Sigma_n/\Sigma_n^{true}$ lower than $\sim 0.5\%$ for $0.1 \le y \le 5$.

\label{lastpage}

\end{document}